\documentclass[fleqn,10pt]{wlscirep}
\usepackage[utf8]{inputenc}
\usepackage[T1]{fontenc}
\usepackage{siunitx}
\title{Dynamic unidirectional anisotropy in cubic FeGe with antisymmetric spin-spin-coupling}

\author[1]{Nicolas Josten}
\author[1]{Thomas Feggeler}
\author[1]{Ralf Meckenstock}
\author[1]{Detlef Spoddig}
\author[1]{Marina Spasova}
\author[2]{Ke Chai}
\author[3]{Iliya Radulov}
\author[2]{Zi-An Li}
\author[3]{Oliver Gutfleisch}
\author[1]{Michael Farle}
\author[1,4,*]{Benjamin Zingsem}
\affil[1]{Faculty of Physics and Center for Nanointegration (CENIDE), University Duisburg Essen, Duisburg, 47057, Germany}
\affil[2]{Institute of Physics, Chinese Academy of Sciences, Beijing 100190, China}
\affil[3]{Department of Material- and Geosciences, Functional Materials, Technische Universit\"at Darmstadt}
\affil[4]{Ernst Ruska Centre for Microscopy and Spectroscopy with Electrons and Peter Gr\"unberg Institute, Forschungszentrum J\"ulich GmbH, 52425 J\"ulich, Germany}

\affil[*]{Benjamin.Zingsem@uni-due.de}

\begin{abstract}
Strong uni\textit{directional} anisotropy in bulk polycrystalline B20 FeGe has been measured by ferromagnetic resonance spectroscopy. Such anisotropy is not present in static magnetometry measurements. B20 FeGe exhibits inherent Dzyaloshinskii-Moriya interaction, resulting in a nonreciprocal spin-wave dispersion. Bulk and micron sized samples were produced and characterized.  By X-band ferromagnetic resonance at $\SI{276}{\kelvin} \pm \SI{1}{\kelvin}$, near the Curie temperature, a distribution of resonance modes was observed in accordance with the cubic anisotropy of FeGe. This distribution exhibits a uni\textit{directional} anisotropy, i.e. shift of the resonance field under field inversion, of $K_{UD}=\SI{960}{\joule\per\metre^{3}} \pm \SI{10}{\joule\per\metre^{3}}$, previously unknown in bulk ferromagnets. Additionally, more than 25 small amplitude standing spin wave modes were observed inside a micron sized FeGe wedge, measured at \SI{293}{\kelvin} $\pm$ \SI{2}{\kelvin}. These modes also exhibit uni\textit{directional} anisotropy. This effect, only dynamically measurable and not detectable in static magnetometry measurements, may open new possibilities for directed spin transport in chiral magnetic systems.
\end{abstract}
\begin{document}

\flushbottom
\maketitle
\thispagestyle{empty}

\section*{Introduction}

Non-centrosymmetric crystal structures, such as the B20 phase of FeGe  \cite{FeGe,FeGe2}, can host chiral spin textures like magnetic skyrmions \cite{doi:10.1002/9783527808465.EMC2016.6263,Muehlbauer915}, which have been proposed as new structures for memory storage applications \cite{Romming636} at room temperature \cite{Woo2016}. Chiral spin structures in general are of significant interest in current magnetic research \cite{Bergmann_2014,PhysRevB.96.094401}. Dzyaloshinsky-Morya-interaction (DMI) \cite{Dzyaloshinsky1958241,PhysRev.120.91} causes a chiral symmetry break of the magnetic interaction and influences the dynamic properties of the magnetic system. For example, the spin wave dispersion becomes non-reciprocal \cite{PhysRevLett.30.125,PhysRevB.88.184404}, as experimentally confirmed by Brillouin spectroscopy \cite{PhysRevLett.114.047201} and an additional phase shift between neighboring spins of a spin wave affects its resonance intensity \cite{1975PhRvL..35.1017P}. The space group P$2_13$ of the FeGe B20 phase has an inherent broken inversion symmetry, but does  not  impose chirality. The chirality, in this case, results from the specific atomic sites occupied by Fe and Ge inside the unit cell \cite{8685788}. The magnetic properties of FeGe were studied using the M\"ossbauer effect \cite{1402-4896-2-4-5-018}, vibrating sample magnetometry \cite{Ludgren_1970} and magnetic susceptibility measurements \cite{LUNDGREN1968175} making FeGe a magnetically well characterized material. \\
In the Heisenberg model of direct nearest neighbour interactions, spin waves (magnons) have a dispersion relation proportional to the wave vector $k$ squared. An antisymmetric contribution to spin-spin interaction results in an additional term in the dispersion relation proportional to $k$ \cite{PhysRevLett.30.125,PhysRevB.88.184404} and therefore a shift with regard to the gamma point. Then spin waves propagating in opposite directions at the same frequency have different wavelengths leading to complex standing waves with a moving phase front. This allows to detect modes, which would cancel and not be detectable in FMR \cite{Zingsem}. \\
 We measured the magnetodynamic properties of a millimeter-sized disk shaped and a micron-sized wedge shaped sample of B20 FeGe using ferromagnetic resonance (FMR) \cite{0034-4885-61-7-001,doi:10.1002/0471266965.com130}. Previous FMR measurements on this material \cite{HARALDSON1972271,HARALDSON1978115,HARALDSON19741237,SMITH1974390} were performed with millimeter sized single crystalline samples. Solving the Landau-Lifshitz-Gilbert equation (LLG) \cite{Landau,1353448} for an FMR like excitation \cite{PhysRevB.96.224407}, we determined magnetic material parameters in the usual way \cite{smit1955}. 

\section*{Sample preparation}
\begin{figure}
\includegraphics[width=\textwidth]{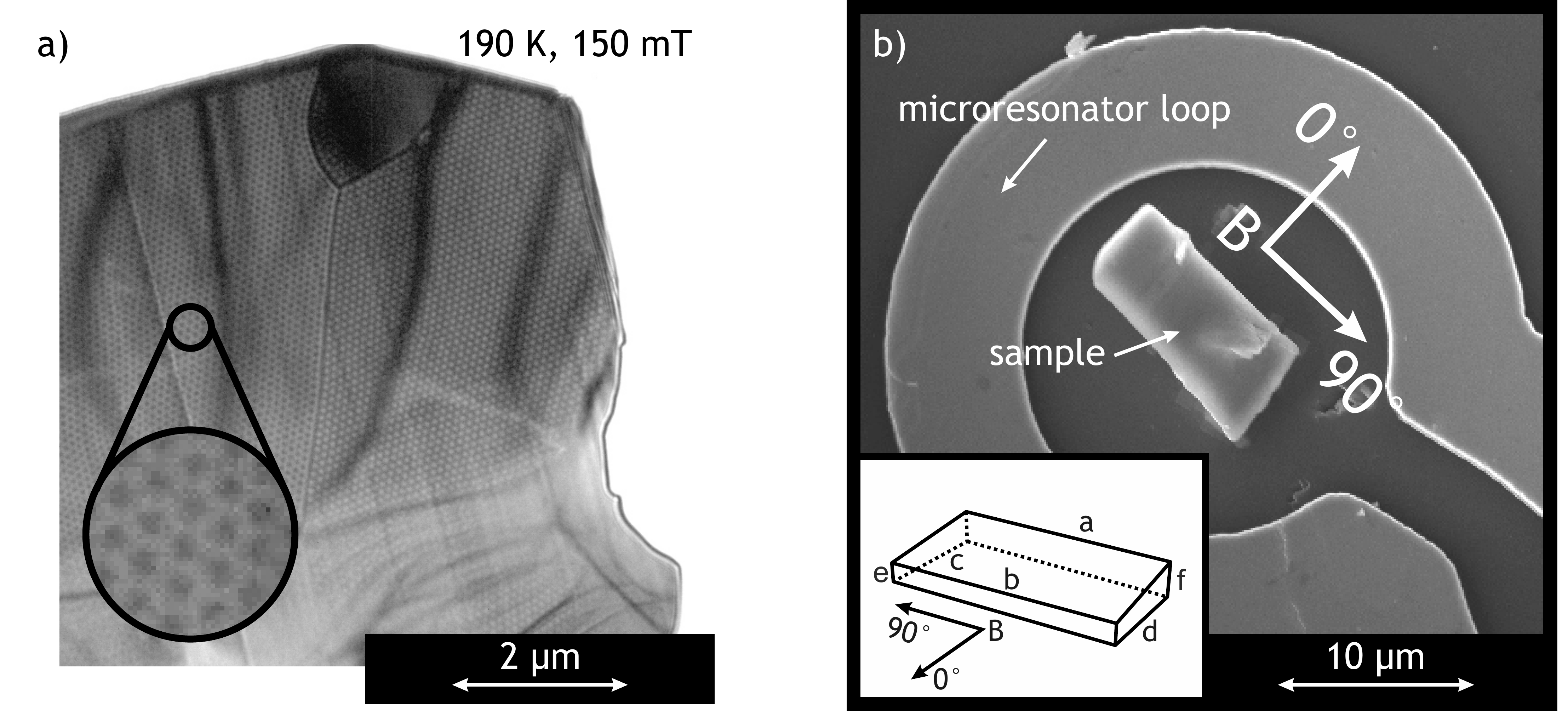}
\caption{(a) Lorentz microscopy image at \SI{190}{\kelvin} and \SI{150}{\milli\tesla} of an FeGe slice cut from the original sample using standard lift-off FIB. The magnetic fields points perpendicular to the sample. The black and white dots represent an ordered skyrmion lattice. (b) Scanning electron micrograph of the specimen inside an R-Type microresonator. The inset shows a schematic representation of the geometry and the directions of the magnetic field B during the experiment. (Dimension of the sample: a~=~11.3~$\pm$~\SI{0.1}{\micro\metre}, b~=~10.9~$\pm$~\SI{0.1}{\micro\metre}, c~=~5.9~$\pm$~\SI{0.1}{\micro\metre}, d~=~5.0~$\pm$~\SI{0.1}{\micro\metre}, e~=~0.9~$\pm$~\SI{0.1}{\micro\metre}, f~=~1.6~$\pm$~\SI{0.1}{\micro\metre}). \label{figure 2}}
\end{figure}

Stoichiometric FeGe was melted, using induction heating and, to guarantee homogeneity, re-melted twice and annealed for \SI{130}{h} at \SI{1000}{K}. Cylinders were formed and a high pressure high temperature synthesis inside a Kawai-type \cite{doi:10.1063/1.1684753} Multianvil Apparatus with Walker-type \cite{Walker} module was applied. This resulted in \SI{95}{\percent} polycrystalline B20 FeGe, confirmed by X-ray diffraction. A maximum of \SI{5}{\percent} of the sample material could consist of secondary phase Iron Germanium. Energy-dispersive X-ray spectroscopy (EDX) measurements also reveal local composition variations with accumulation of iron (Fe:Ge 55:45). Further investigations with Lorentz microscopy show the formation of helices and skyrmions (fig. \ref{figure 2} (a)) in accordance to \cite{article50}. Micron sized samples (figure \ref{figure 2} (b)) with wedge shaped geometries were cut using a Focused Ion Beam (FIB -– FEI Helios nanolab 600) and placed inside an R-Type microresonator \cite{0957-4484-22-29-295713,doi:10.1063/1.2964926,NARKOWICZ2005275} using standard lift-off FIB (Omniprobe manipulator with Pt gas insertion system) technique. During the lift-off process a carbon coating with up to \SI{100}{\nano\metre} thickness and up to \SI{15}{\percent} platinum contamination \cite{0957-4484-17-15-028} could not be avoided. Furthermore, the lift-off process used Gallium as cutting ions and resulted in a localized deposition of a maximum of \SI{2.6}{\percent} of Ga (as verified by EDX).

\begin{figure}
\includegraphics[width=0.85\textwidth]{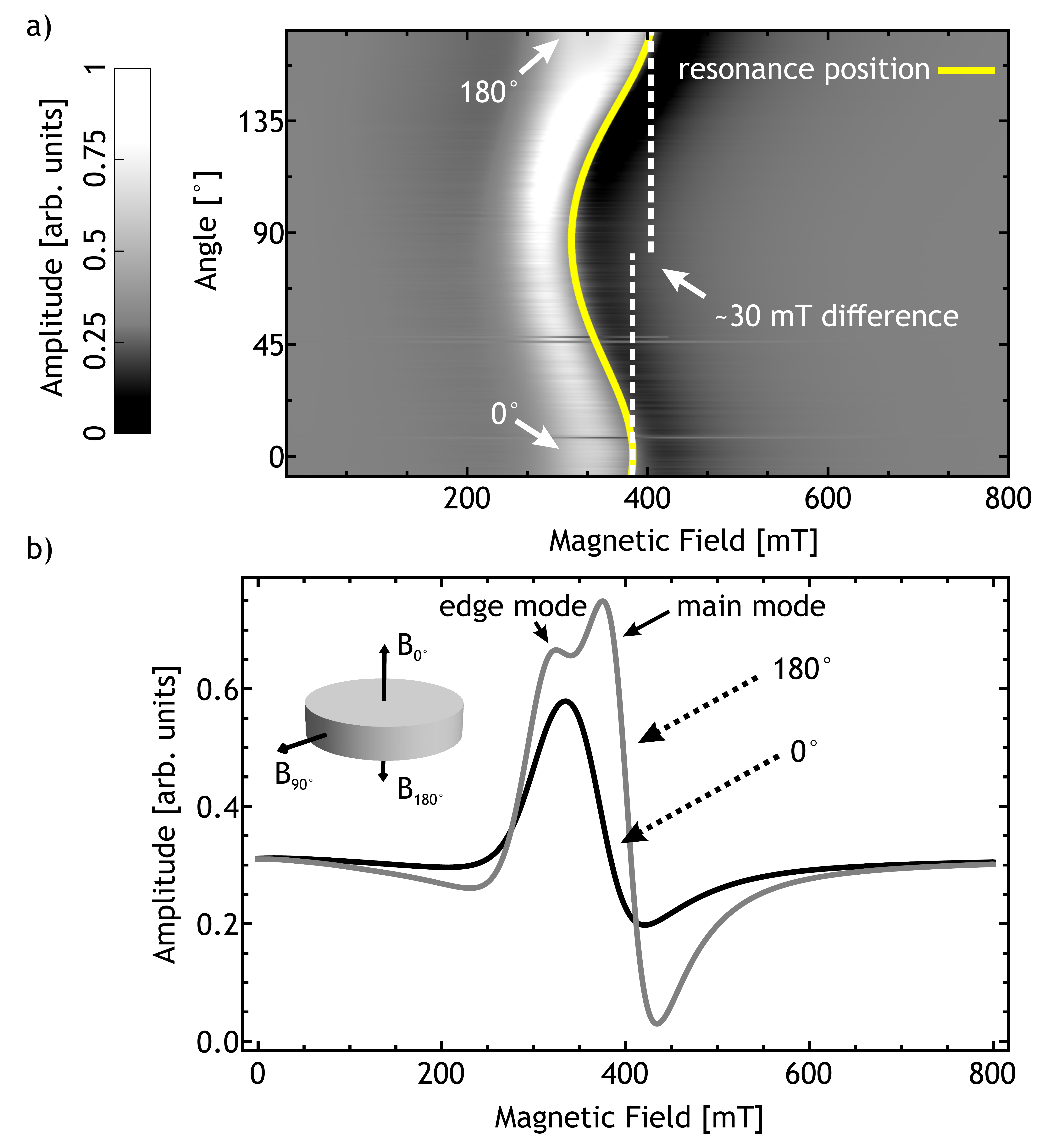}
\caption{(a) Angular dependent out-of-plane (differentiated) FMR spectra shown as an amplitude contour plot at \SI{276}{\kelvin} $\pm$ \SI{1}{\kelvin} and f$_{\text{Microwave}}$ = \SI{9.517}{\giga\hertz} $\pm$ \SI{0.006}{\giga\hertz}. The yellow line marks the angular dependent resonance field position. The dotted white lines mark the position of the hard direction at $\ang{0}$ and $\ang{180}$. They have been extended to the middle of the figure for better comparison of the \SI{30}{\milli\tesla} field difference. (b) shows the spectra of the same FMR measurement at $\theta=\ang{0}$ and $\theta=\ang{180}$. The resonance spectra at $\ang{180}$ appears to consist of two resonance lines. This is due to edge resonances inside the sample \cite{David,HARALDSON19741237}. Additionally a schematic representation of the sample can be seen with the most important field positions marked. \label{figure 3}}
\end{figure}

\begin{figure}
\includegraphics[width=\textwidth]{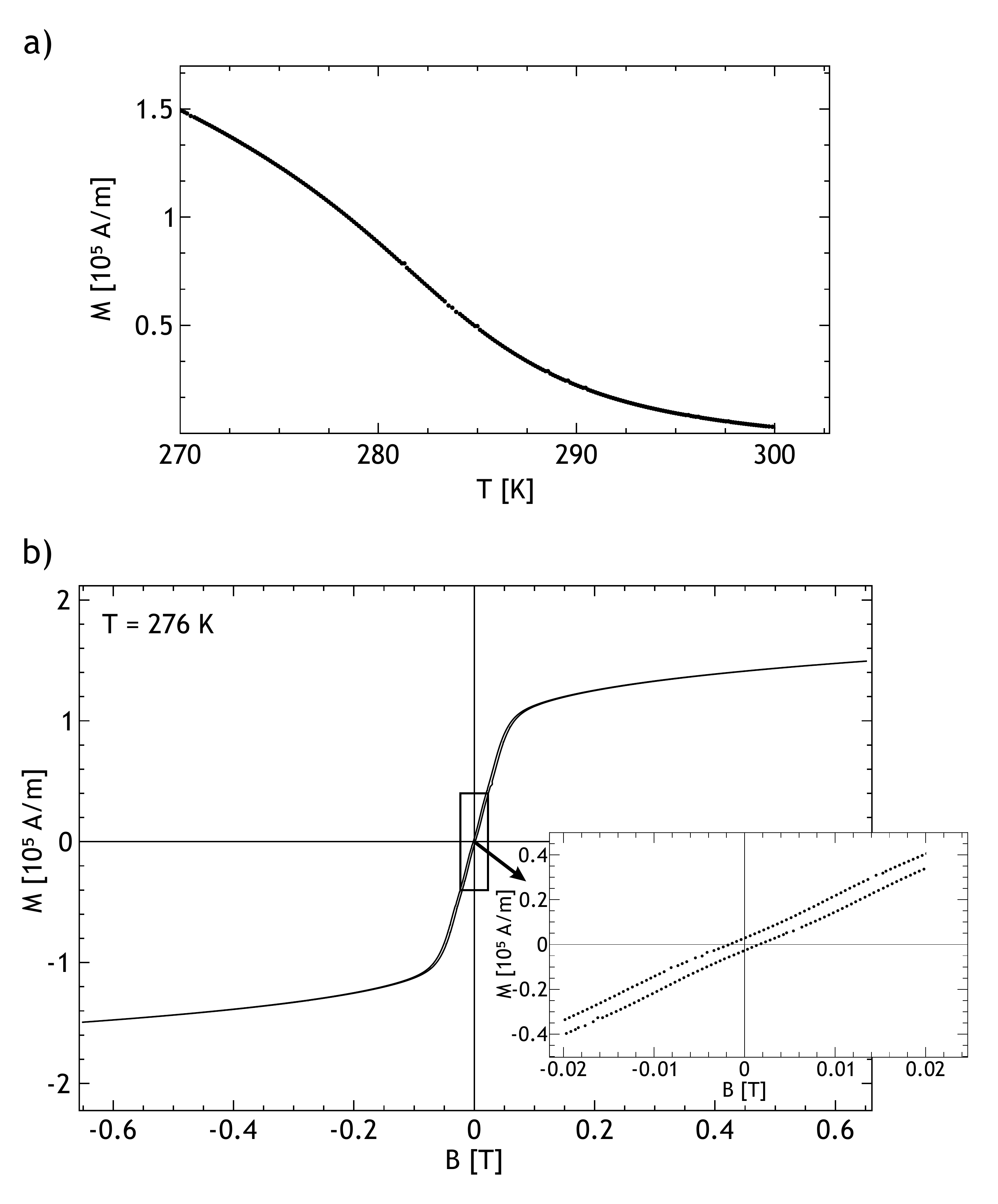}
\caption{(a) Temperature dependent measurement of the magnetization at \SI{310}{mT} external field.  A part of the original sample was used for the measurement. (b) Hysteresis loop measured by vibrating sample magnetometry at \SI{276}{K}. The sample is the same as in (a). The Magnetisation M is plotted against the magnetic field B. The hysteresis shows no asymmetry or exchange bias.  \label{figure H}}
\end{figure}

\begin{figure}
\includegraphics[width=\textwidth]{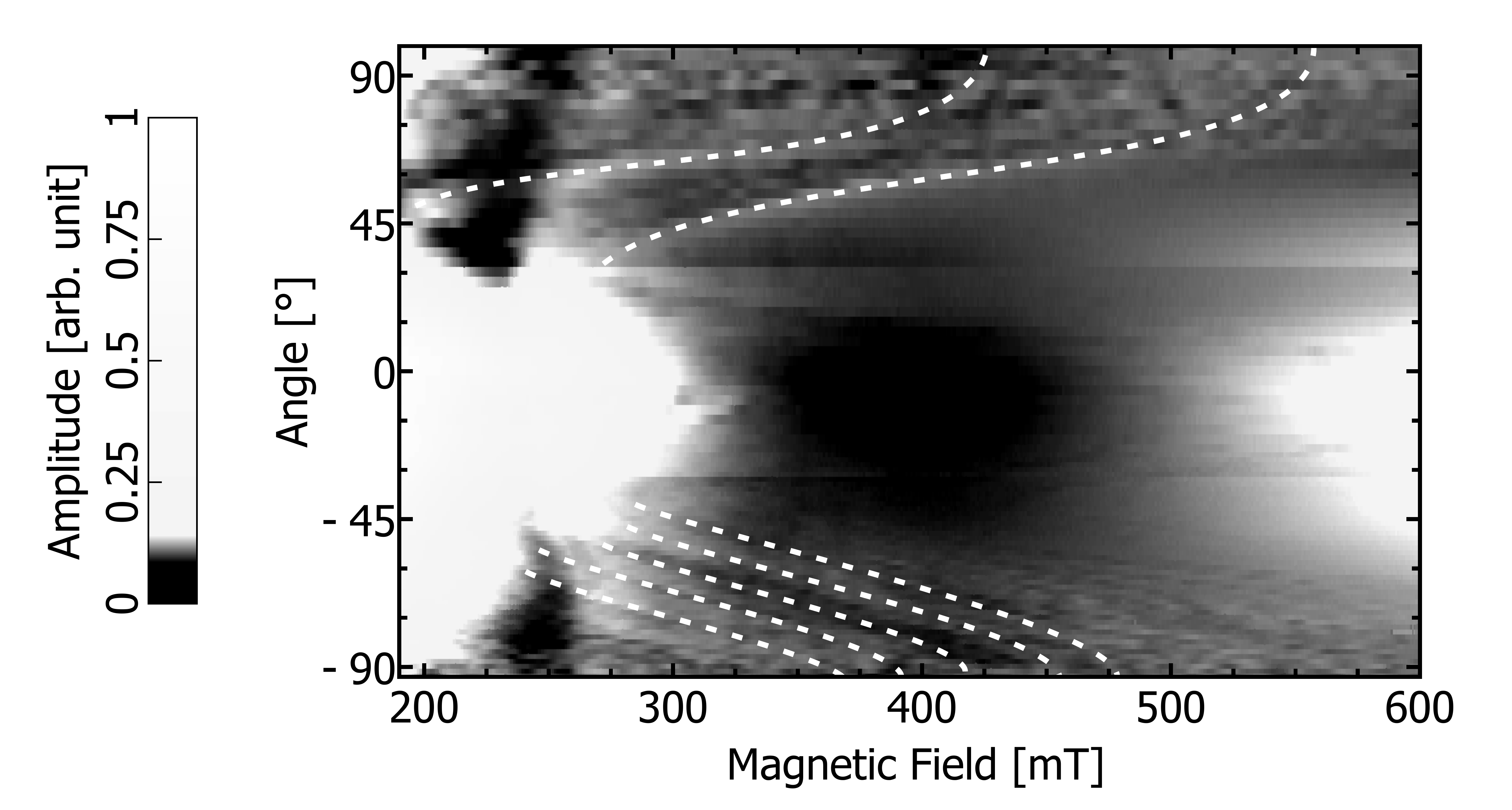}
\caption{A grey scale contour plot of the (differentiated) FMR signal amplitude of the micron sized wedge shaped sample as a function of applied magnetic field for different orientations of the magnetic field between \ang{-93} and \ang{99} (compare fig. \ref{figure 2} (b)) at \SI{9.134}{\giga\hertz} $\pm$ \SI{0.006}{\giga\hertz}. The scale bar is depicted on the left. The dotted white lines indicate angular dependent resonance fields of spin wave modes. \label{figure 4}}
\end{figure}

\begin{figure}
\includegraphics[width=\textwidth]{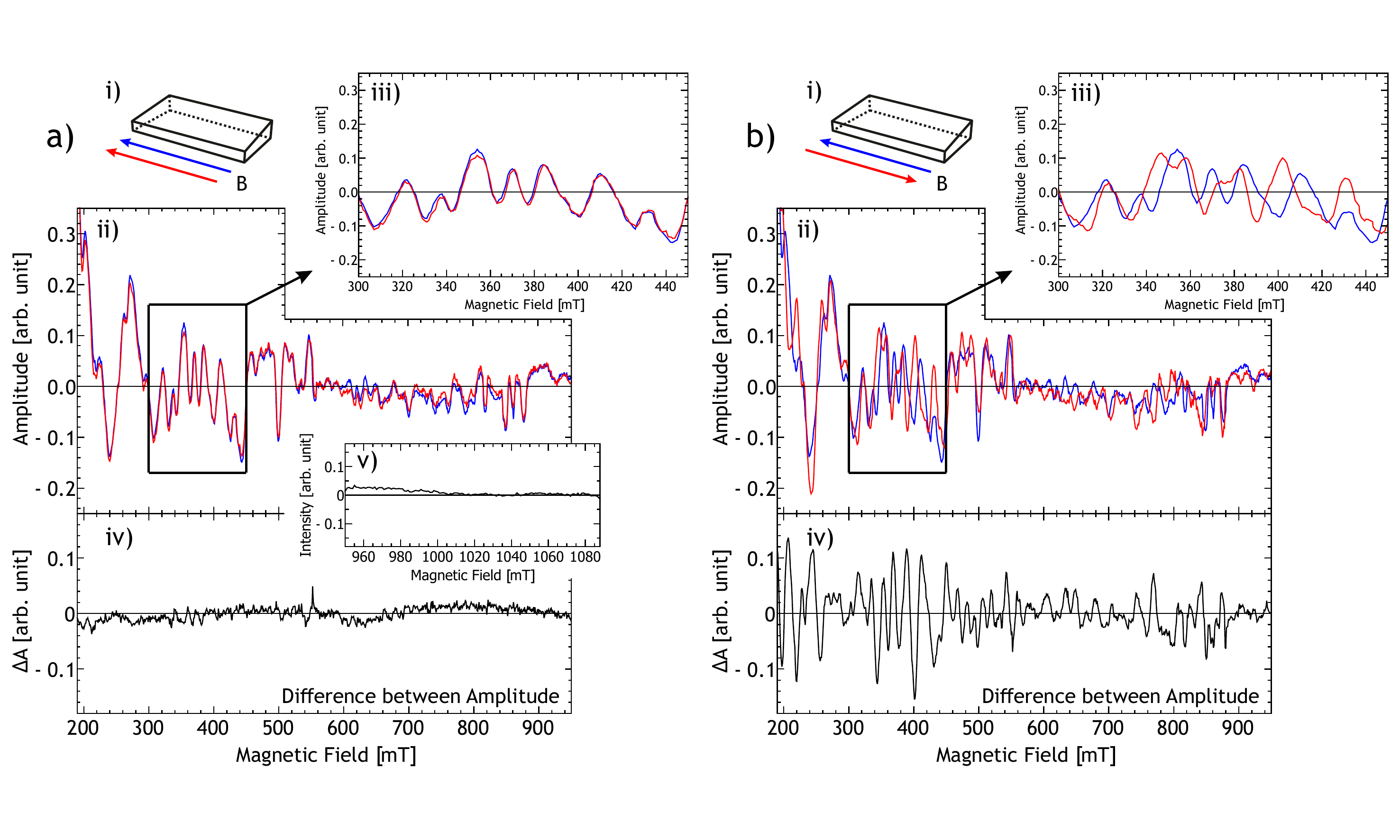}
\caption{Analysis of the bidirectional (differentiated) FMR measurements ($\sim \, \SI{293}{\kelvin}$) at \ang{81} and \SI{3.587}{\giga\hertz} $\pm$ \SI{0.006}{\giga\hertz}, of the specimen shown in fig. \ref{figure 2} (b). (a,ii) and (b,ii) comparison between two different (differentiated) FMR measurements, the former with the same magnetic field direction, the latter at opposite field directions. (a,i) and (b,i) show a schematic representation of the sample and the respective magnetic field directions of the compared measurements. (a,iii) and (b,iii) depict the highlighted areas in (a,ii) and (b,ii) in detail. (a,iv) and (b,iv) are the plotted differences between the compared measurements. (a,v) noise floor of the measurement in a magnetic field region without resonances. \label{figure 5}}
\end{figure}

\section*{Experimental}
FMR spectra of a bulk polycrystalline, nearly disc shaped piece of FeGe with a diameter of \SI{3.78}{\milli\metre} and a thickness of \SI{0.78}{\milli\metre} (\ref{figure 3}(b) inset) was acquired in a range of \SI{800}{\milli\tesla} to \SI{0}{\milli\tesla} at a frequency of \SI{9.517}{\giga\hertz} $\pm$ \SI{0.006}{\giga\hertz}. The field was applied at angles of \ang{0} to \ang{180} in steps of \ang{0.5} from out-of-plane to in-plane and to the opposite out-of-plane orientation. The measurement can be seen in fig.\ref{figure 3} (a) shown as an amplitude contour plot. The temperature is \SI{276}{\kelvin} $\pm$ \SI{1}{\kelvin}, which is below the Curie Temperature of T$_\text{C}=\SI{280}{\kelvin}$ \cite{LUNDGREN1968175}, where the sample is ferromagnetic \cite{article50}. The angular precision of our experimental setup is better than $\ang{0.05}$ and the precision of the magnetic field is better than \SI{0.5}{\milli\tesla} with a relative precision of \SI{0.005}{\milli\tesla}. \\
Resonance lines in the FMR spectra are identified by a successive local maximum and minimum amplitude. We observe a distribution of resonances, which is in agreement to previous FMR investigations \cite{HARALDSON1972271} of single crystalline FeGe. Each crystallite in the sample is contributing to this resonance distribution. They are all influenced by the applied external field and the demagnetization field in the sample, due to its general shape. However, their resonance fields vary with respect to the applied magnetic field due to the different symmetry axis of the cubic anisotropy in each crystallite. We simulated the resonance distribution using the known magnetocrystalline anisotropy of FeGe \cite{HARALDSON1972271} and a random orientation of crystallites and compared it to the measurement. This can be found in the supplementary sec. S1. Figure \ref{figure 3} (a) shows the differentiated angular dependent FMR spectra as a grey scale contour plot. The out-of-plane orientations are depicted in detail in fig. \ref{figure 3} (b). The resonance line exhibits a unidirectional anisotropy, indicated by a difference in the positon of maximum microwave absorption comparing \ang{0} and \ang{180}. A similar anisotropy is observed in systems with exchange-bias \cite{PhysRevB.38.6847}. Hence we performed additional magnetometry measurements, to exclude the presence of exchange bias in our system (fig. \ref{figure H}). No such anisotropic behaviour is observed in static magnetometry using vibrating sample magnetometry (VSM). We therefore conclude, that this anisotropy must be dynamically induced under resonant excitation. Note that it cannot be equated with the linear contribution to the spin wave dispersion, as this changes directionality in accordance with the magnetic field direction. Due to the skin depth of approximately \SI{e-3}{\milli\metre} \cite{HARALDSON1972271} one must, to fully reproduce the FMR lineshape, solve the non-uniform LLG \cite{HARALDSON19741237} taking the shape of the sample into account. However, we show exemplary in the supplementary sec. S1 that a Dysonian lineshape \cite{lineshape,dyson2} and the known magnetocrystalline anisotropy of FeGe \cite{Ludgren_1970} are able to reproduce the measured FMR lineshape satisfyingly, which is sufficient for our needs. The position of resonance was obtained by subtracting the background and locating the zero crossing of the resonance line. We analyzed the angular dependent spectra using eq. \ref{equation 3} as a model for the free energy density F. To account for the observed unidirectional symmetry in the angular dependent resonance field position, an additional unidiretional field contribution needs to be introduced. In this model an additional anisotropy field $B_U = K_{UD}/M$ is used. This unidirectional contribution is merely a descriptive model to account for the observed phenomenon. It cannot be seen as an additional magnetocristalline anisotropy but rather as an emergent symmetry contribution which arises under dynamic excitation. In the supplementary sec. S2 the shape of such a unidirectional free energy density is shown. Additionally, a demagnetization and Zeeman term are considered.

\begin{equation}
F = -K_{UD} \cdot \text{cos}\left( \theta \right) + \frac{\mu_0}{2} \cdot \vec{M} \cdot N \cdot \vec{M} - \vec{M} \cdot \vec{B}
\label{equation 3}
\end{equation}

The demagnetisation tensor $N$ ($N_{zz}=0.676,N_{xx,yy}=0.162$) was deduced, using the demagnetisation tensor of a cylinder as described in \cite{0022-3727-39-5-001}. The known g-factor of FeGe (g = 2.07) \cite{HARALDSON1978115} was used. $\theta$ is the out-of-plane angle of the magnetisation $M$ and $B$ the external magnetic field. Additionally, the magnetisation $M$ is considered as a fit parameter. The obtained parameters are $K_{UD}=\SI{960}{\joule\per\metre^{3}} \pm \SI{10}{\joule\per\metre^{3}}$, $M=\SI{82580}{\ampere\per\metre} \pm \SI{200}{\ampere\per\metre}$. The magnetization matches the magnetization measured by VSM at \SI{281}{\kelvin}, \SI{5}{\kelvin} above the temperature measured by a sensor below the sample. This offset is likely due to microwave heating. Following \cite{HARALDSON1978115}, we assume this to be the uniform FMR mode of cubic FeGe. \\
Figure \ref{figure 4} shows the angular dependent FMR spectra (\SI{293}{\kelvin} $\pm$ \SI{2}{\kelvin}, $f_{\text{Microwave}}$= \SI{9.134}{\giga\hertz} $\pm$ \SI{0.006}{\giga\hertz}) of the wedge shaped FeGe sample (fig. \ref{figure 2} (b)) measured inside a microresonator as a grey scale contour plot. Multiple resonances are visible in the spectra, which exhibit anisotropic behavior. The anisotropy is directed such that the resonance field increases when the static field is applied parallel to the long (dipolar-easy) axis of the sample. This suggests that these modes are spinwaves with energies below that of the gamma point (FMR mode), which may be induced by strong dipolar coupling \cite{RevModPhys.30.1} or DMI. Around $\pm$ \ang{90}, the number of superimposing resonances and the complex mode intesity distribution \cite{Zingsem} make it difficult to separate individual lines. We assume that these resonances arise due to geometrical confinement of the modes in our specimen (fig. \ref{figure 2}). Consequently, the inclined surface of our wedge results in different geometrical boundary conditions at the same time. Bidirectional measurements along the $\pm$~\ang{81} direction as shown in fig. \ref{figure 5}, however, reveal a clear unidirectional shift of the resonances under field reversal. Figure \ref{figure 5} (a) shows the reproducibility of resonances for field sweep up and field sweep down, whereas fig. \ref{figure 5} (b) illustrates that under field reversal, the resonance position of the spinwaves has shifted. Hence, we find a unidirectional anisotropy. Figure \ref{figure 5} (c) shows the noise floor of our spectrometer in a field region where no resonances are observed.

\section*{Conclusion}

From angular dependent ferromagnetic resonance we find an unexpected dynamic unidirectional anisotropy (fig. \ref{figure 3}) in the magnetic excitation of FeGe just below the Curie temperature. This anisotropy is of a dynamic character, since it is not detectable in static hysteresis measurements. The magnitude of the unidirectional anisotropy of the bulk resonance line is $K_{UD}=\SI{960}{\joule\per\metre^{3}} \pm \SI{10}{\joule\per\metre^{3}}$. Spin waves, detected at \SI{293}{\kelvin} $\pm$ \SI{2}{\kelvin} for sample sizes with micrometer dimensions, also exhibit unidirectional anisotropy (fig. \ref{figure 5}).

\section*{Methods}
A conventional Bruker X-band FMR spectrometer was used for FMR measurements on the millimetre sized FeGe sample (see fig. \ref{figure 3}) inside a cylindrical TE$_{011}$ cavity.  
FMR measurements on the micron sized FeGe sample (figure \ref{figure 2} (b), fig. \ref{figure 4} and fig. \ref{figure 5}) were performed inside an R-Type microresonator \cite{0957-4484-22-29-295713,doi:10.1063/1.2964926,NARKOWICZ2005275}. The resonator was connected to a Varian E102 microwave bridge. The modulated microwave reflection was recovered using a SRS SR830DSP lock-in amplifier.

\bibliography{literature}

\section*{Acknowledgements}

We like to thank Norimasa Nishiyama and Shrikant Bhat for the support at the sample synthesis and Igor Barsukov and Konstantin Skokov for fruitful discussions. 

B Z acknowledges that the research leading to these results has received funding from the European Research Council under the European Union's Seventh Framework Programme (FP7/2007-2013)/ ERC grant agreement number 320832.

T F acknowledges financial support by German Research Foundation (DFG project: OL513/1-1) and the Austrian Science Fund (FWF project: I 3050-N36).

I R and O G acknowledge financial support by the German federal state of Hessen through its excellence program LOEWE "RESPONSE".

\section*{Author contributions statement}

N J and T F performed magnetic resonance experiments with support from B Z and R M. Z L and K C performed electron microscopy.  I R and D S prepared the samples and perfomed analytic characterizations with help from O G. M S performed magnetometry.
B Z and T F concieved and planned the experiments. B Z and N J analysed the data and wrote the manuscript with support from T F and  M F.
B Z encouraged N J to investigate the presented samples and supervised the findings of this work.
All authors discussed the results and contributed to the final manuscript. B Z supervised the project.

\section*{Additional information}
\textbf{There are no competing interests} 

\end{document}